\documentclass[12pt, manuscript]{aastex}

\usepackage{natbib, graphicx, subfigure}
\begin{document}  

\title{Imaging Carbon Monoxide Emission in the Starburst Galaxy NGC~6000}  
\author{Sergio Mart\'{\i}n}  
\affil{Harvard-Smithsonian Center for Astrophysics, 60 Garden St.,  02138, Cambridge, MA, USA}
\email{smartin@cfa.harvard.edu} 
\author{Matthew R. George}  
\affil{Department of Astronomy, University of California, Berkeley, CA 94720, USA}
\affil{Harvard-Smithsonian Center for Astrophysics, 60 Garden St.,  02138, Cambridge, MA, USA}
%\email{mrgeorge@post.harvard.edu} 
\author{David J. Wilner}  
\affil{Harvard-Smithsonian Center for Astrophysics, 60 Garden St.,  02138, Cambridge, MA, USA}
\author{Daniel Espada}  
\affil{Harvard-Smithsonian Center for Astrophysics, 60 Garden St.,  02138, Cambridge, MA, USA}
\affil{Instituto de Astrof\'isica de Andaluc\'ia, CSIC, Granada, Spain}

\begin{abstract}
We present measurements of carbon monoxide emission in the central region of the nearby starburst NGC~6000 taken with the Submillimeter Array.
The $J=2-1$ transition of $^{12}$CO, $^{13}$CO, and C$^{18}$O were imaged at a resolution of $\sim3''\times2''$ ($450\times300$\,pc).
We accurately determine the dynamical center of NGC\,6000 at $\alpha_{J2000.0}=15^h49^m49\fs5$ and $\delta_{J2000.0}=-29\arcdeg23\arcmin13''$
which agrees with the peak of molecular emission position.
The observed CO dynamics could be explained in the context of the presence of a bar potential affecting the molecular material,
likely responsible for the strong nuclear concentration where more than $85\%$ of the gas is located.
We detect a kinematically detached component of dense molecular gas at relatively high velocity which might be fueling the star formation.
A total nuclear dynamical mass of $7\times10^9\,M_\odot$ is derived and a total mass of gas of $4.6\times10^8M_\odot$, yielding
a $M_{gas}/M_{dyn}\sim6\%$, similar to other previously studied barred galaxies with central starbursts.
We determined the mass of molecular gas with the optically thin isotopologue C$^{18}$O and we estimate a CO-to-H2 conversion factor
$X_{CO}=0.4\times10^{19}\,\rm cm^{-2}(K\,km\,s^{-1})^{-1}$ in agreement with that determined in other starburst galaxies.
%We estimated isotopic ratios of $\rm^{12}C/^{13}C=20\pm1$ and $\rm^{16}O/^{18}O=76\pm14$, a factor of two lower than the values derived
%for other starburst galaxies.
\end{abstract}

\keywords{galaxies: fundamental parameters --- galaxies: ISM --- galaxies: kinematics and dynamics --- galaxies: nuclei --- galaxies: starburst --- radio lines: galaxies}

\section{Introduction}

The relation between the morphology and kinematics of the gas in the inner region of galaxies and the connection to its nuclear activity is
not completely understood.
%The detailed mechanics of starbursts, bars, and other nuclear activity in spiral galaxies are not completely understood. 
Nuclear regions are often obscured by dust at optical wavelengths but can be viewed in continuum infrared dust emission as well as molecular line
emission. CO can be used as a tracer of molecular gas, and its rotational transitions are observed at millimeter and sub-millimeter wavelengths.
Observations of the morphology and kinematics of the molecular gas content in the central regions of spirals can provide insight into the dynamical processes occurring there 
\citep[e.g.,][]{jogee05, perez00}. Studies indicate that instabilities in bars and dissipation of gas clouds remove angular momentum from material 
orbiting the galactic center, driving gas and dust inward to fuel star formation \citep[e.g.,][]{pfenniger90,Sakamoto99b,knapen02,jogee05,Sheth05}.

NGC\,6000 is a nearby \citep[D$\sim$31.6\,Mpc,][]{pizzella05} barred spiral starburst galaxy.
Despite its proximity and brightness, this galaxy has not been studied in as great detail as others of similar distance and 
luminosity, due to its southern declination.
Hubble Space Telescope observations reveal several bright sources in circumnuclear 
star-forming rings and a large-scale ($\gtrsim$~1~kpc) bar \citep{carollo97, carollo02, fathi03}.
However,
%other analysis of the same images have not detected the existence of a bar because of irregularities in the
no such barred structures are identified towards the more irregular
nuclear region \citep{carollo99, hunt04}.
Infrared observations towards the nucleus of NGC~6000 have found significant polycyclic aromatic hydrocarbon (PAH) emission \citep{siebenmorgen04} and determined a 
dust temperature of $29.2\pm2.6$~K \citep{yang07}.
NGC\,6000 appears bright in CO emission \citep{young95, mauersberger99} as well 
as \ion{H}{1} \citep{koribalski04} as detected with single-dish telescopes.
However no detailed morphological and kinematical study of the gas content has been performed yet.
A compilation of the observed and derived properties for NGC\,6000 are presented in Table~\ref{galaxyparams}.
%Modern interferometric arrays have allowed for arcsecond imaging of galactic structures, greatly improving our ability to analyze morphology at 
%smaller scales.
%This galaxy is a nearby spiral starburst that has not been studied in as great detail as others of similar distance and 
%luminosity, due to its southern declination. 

In this paper we present observations of the carbon monoxide emission in the $J=2-1$ transition of the three brighter isotopologues
($\rm^{12}CO$,$\rm^{13}CO$,$\rm C^{18}O$) towards \object{NGC\,6000} using the Submillimeter Array \citep[SMA;][]{ho04}.
This is the first high resolution morphological study of the molecular component towards NGC\,6000.

%Aside from gaining morphological information for NGC~6000, we derive gas number densities and kinematic mass estimates. 
%We also use the relative intensities of the detected CO isotopes, $^{12}$CO, $^{13}$CO, and C$^{18}$O, to determine the optical depth and 
%abundance ratios, following \citet{goldsmith99}. In the next section we will describe our observations and data reduction. 
%The following sections describe the data analysis and derived properties of the galaxy, concluding with a summary of our results. 
%We use a distance of 30.1~Mpc (check?) for calculations \citep{tully88}.

%%%NGC 6000 parameters table%%%
%apj
%\begin{deluxetable*}{lcc}

%manuscript

%%%

\section{Observations}
\label{Observations}

Observations were carried out on 2007 February 20 with the Submillimeter Array (SMA) atop Mauna Kea, Hawaii.
Seven out of the eight antennas of the array were operational and placed in the ``compact-north" configuration, with baselines
ranging $16-120$\,m.
The SIS receivers were tuned to the CO $J=2-1$ frequency (230.538\,GHz) in the upper side band (USB), redshifted to the NGC\,6000
average systemic velocity of 2145\,km\,s$^{-1}$ \citep{albrecht07}.
% WHY THIS VELOCITY???
% of 2141\,km\,s$^{-1}$.
The correlator was configured so that the $J=2-1$ transitions of the isotopic substitutions $^{13}$CO (220.398\,GHz) and
C$^{18}$O (219.560\,GHz) were simultaneously observed in the lower side band (LSB).
Spectral resolution was 0.8125\,MHz ($\sim1\rm\,km\,s^{-1}$).
System temperatures ranged from $125-250$\,K, with a zenith opacity of $0.1-0.15$ as measured at 225\,GHz from CSO.

A single field (field of view $\sim55''$ at 230\,GHz) was observed towards the central position $\alpha_{J2000.0}=15^h49^m49\fs40$ and 
$\delta_{J2000.0}=-29\arcdeg23\arcmin11\farcs0$ as measured by IRAS \citep{sanders03}.
We achieved a resolution of $2.8''\times1.7''$ in the uniform weighted maps and $3.7''\times2.2''$ in the natural weighted maps.
At the distance of NGC\,6000, $1''$ is equivalent to a projected distance of $\sim153$\,pc.
Passband calibration was derived from spectra of Saturn and 3C~279.
Absolute flux calibration (within a $10\%$ accuracy) was determined from observations of Callisto.
Gain fluctuations were corrected with the nearby quasar J1517-2422
%1517-243 
at a distance of $8.7^\circ$ from NGC\,6000.
The applied gain corrections were checked by imaging another nearby quasar, J1625-2527,
%1625-254 
at a distance of $8.9^\circ$.
%as secondary gain calibrator.
Data calibration and reduction was performed using the \textsc{MIR-IDL} package and imaging using \textsc{MIRIAD} \citep{sault95}.
% \citep{sault95}.

\section{Results}
\label{Results}

\subsection{Continuum Emission}
%theoretical rms noise in map for 9072 visibilities is     3.37 mJy/beam
% 3/4 of the band free of emission => rms= 3.9 mJy
% At these frequency: Jy/K = 0.34
% TOTAL RMS ~ 4.6mJy
Continuum emission from the nucleus of NGC\,6000 was imaged using the CO emission free channels in the USB.
The spectral energy distribution (SED) of NGC\,6000 shows how most of its continuum emission at 1.3\,mm stems from thermal dust emission
with no significant evidence of free-free contribution \citep{roche93}.
In Fig.~\ref{fig:composite} ({\it left panel}) we show the 1.3\,mm continuum emission over the an average image of the J, H, and K 2MASS bands \citep{Skrutskie06}.
The position of the peak of emission at
$\alpha_{J2000.0}=15^{\rm h}49^{\rm m}49\fs5,\,\delta_{J2000.0}=-29^\circ23'13''$
matches that of 2MASS emission within the resolution.
This position also matches within $2''$ that of the FIR from low resolution IRAS images \citep{sanders03} and within $1''$ that of the VLA 1.4\,GHz
peak \citep[$18''$ resolution,][]{Condon96}.
Together with the CO mapping presented in Sect.~\ref{sect.COemission}, we have obtained
the most accurate determination of the nucleus position in NGC\,6000, with a resolution better than the $2.5''$ of 2MASS \citep{Skrutskie06}.

The faint unresolved continuum source has a total flux of $13\pm3$\,mJy as measured from our observations.
This flux seems to agree with the $18\pm8$\,mJy flux observed at 1.3\,mm with the $19.5''$ beam of the JCMT telescope \citep{roche93}.
However, the JCMT flux might be overstimated given that the contribution of the CO line (Sect.~\ref{sect.COemission}) to the continuum in the 64\,GHz bolometer band
would be $\sim13$\,mJy. Therefore, the corrected JCMT continuum flux would be a factor of two lower than our measurement.
On the other hand, the CO emission would contribute $\sim16$\,mJy to the $42.7\pm3.9$\,mJy average flux measured by the $11''$ beam of the IRAM 30\,m
over its 50\,GHz bolometer band \citep{chini95}.
This corrected flux is a factor of two larger than what we recover from our observations.
An even larger 1.3\,mm flux of $60.6\pm7.8$\,mJy was measured by SEST telescope, which is likely due to its larger beam ($24''$) collecting most of
the extended emission \citep{chini95}.
Indeed, the extended 1.3\,mm continuum emission is estimated to be $98$\,mJy (CO corrected) in the inner $70''$ with the 5 pointing map observations with the
SEST telescope \citep{albrecht07}.
Assuming the continuum measurements from SEST, we only recovered the compact $\sim15\%$ of the continuuum emission in our maps.

\subsection{Molecular Gas Emission}
\label{sect.COemission}

\subsubsection{Comparison to single dish}
% Our position:   15:49:49.5  -29:23:13
% Chini position: 15:49:49.36 -29:23:13 15:46:44.1 -29:14:08
The total integrated flux recovered from the $^{12}$CO $2-1$ image is $\rm 1080\pm60\,Jy\,km\,s^{-1}$.
Single dish observations of this transition with the $24''$ beam of the SEST telescope resulted in a total detected flux of $780\pm80$\,Jy\,km\,s$^{-1}$
\citep{chini96}, where we assumed a conversion factor of $\sim25$\,Jy/K.
This difference is contrary to the expected lower flux in the interferometric maps due to filtering of extended emission.
Indeed, the $\sim 30\%$ flux difference is too large to be attributed to absolute flux calibration error of both SEST and SMA data.
Although SEST observations were taken towards a nominal position $2''$ away from the position of peak emission derived from our map,
this difference would only explain a $<4\%$ lower measured flux.
%, so calibration errors are likely to explain
%the inconsistency between the results.
In order to understand this measured flux, we convolved our map to a $24''$ resolution.
The asymmetric lineshape in the data from \citet{chini96} clearly differs from that obtained at the center of the convolved map and significantly
matches the lineshape at a position $\gtrsim5''$ south of the central position, consistent with the SEST pointing accuracy.
This difference in the observing position easily accounts for a difference in the integrated intensity of $>20\%$ with respect to the central position.
Though the uncertain comparison with single-dish data does not allow us to accurately determine the missed flux in our maps, we can assume that
no significant amount of flux have been filtered out and that most of the extended CO emission has been recovered in our maps.

\subsubsection{$^{12}$CO morphology}
\begin{figure}
\centering
\includegraphics[angle=0,width=0.4\textwidth]{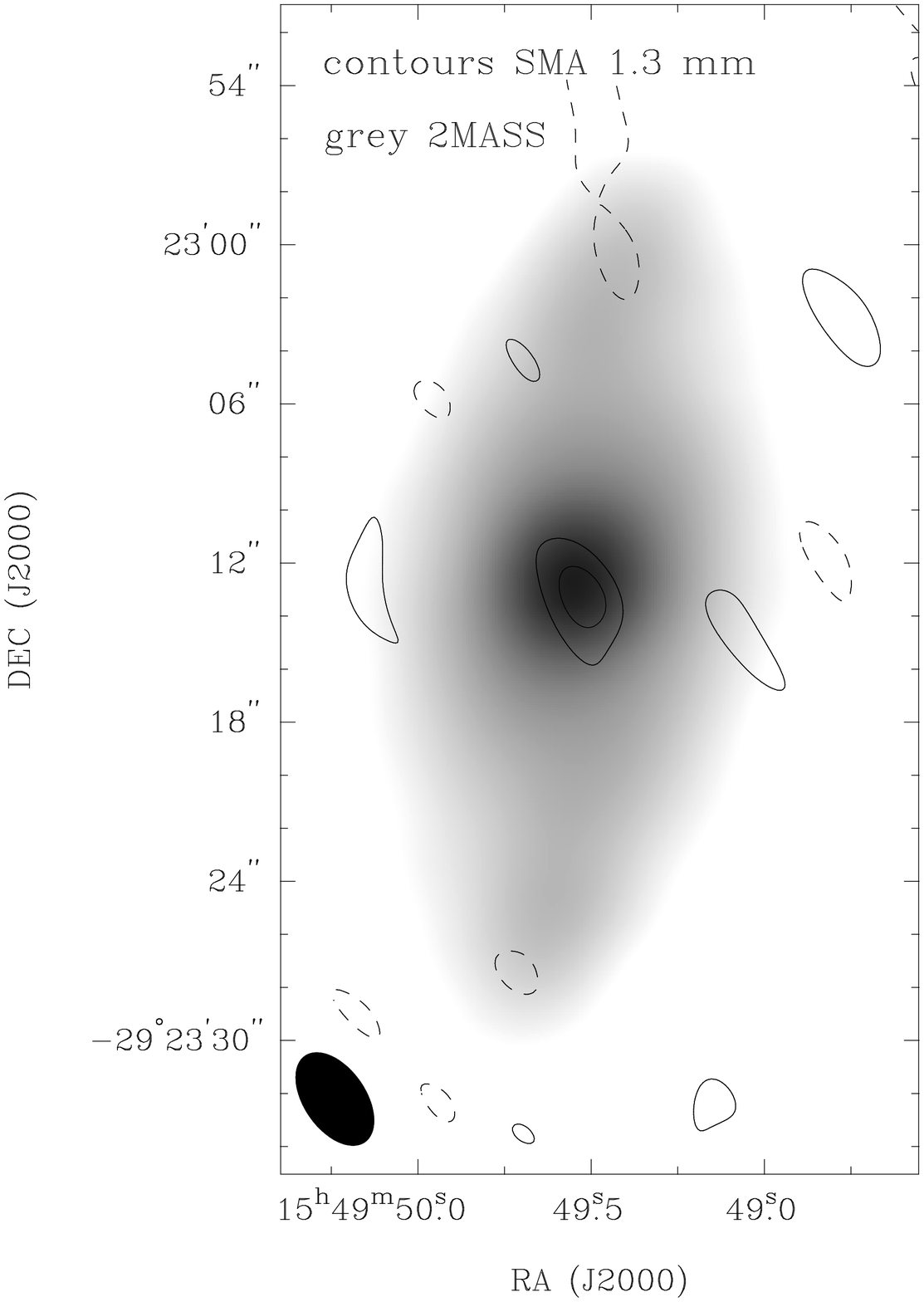}
\includegraphics[angle=0,width=0.4\textwidth]{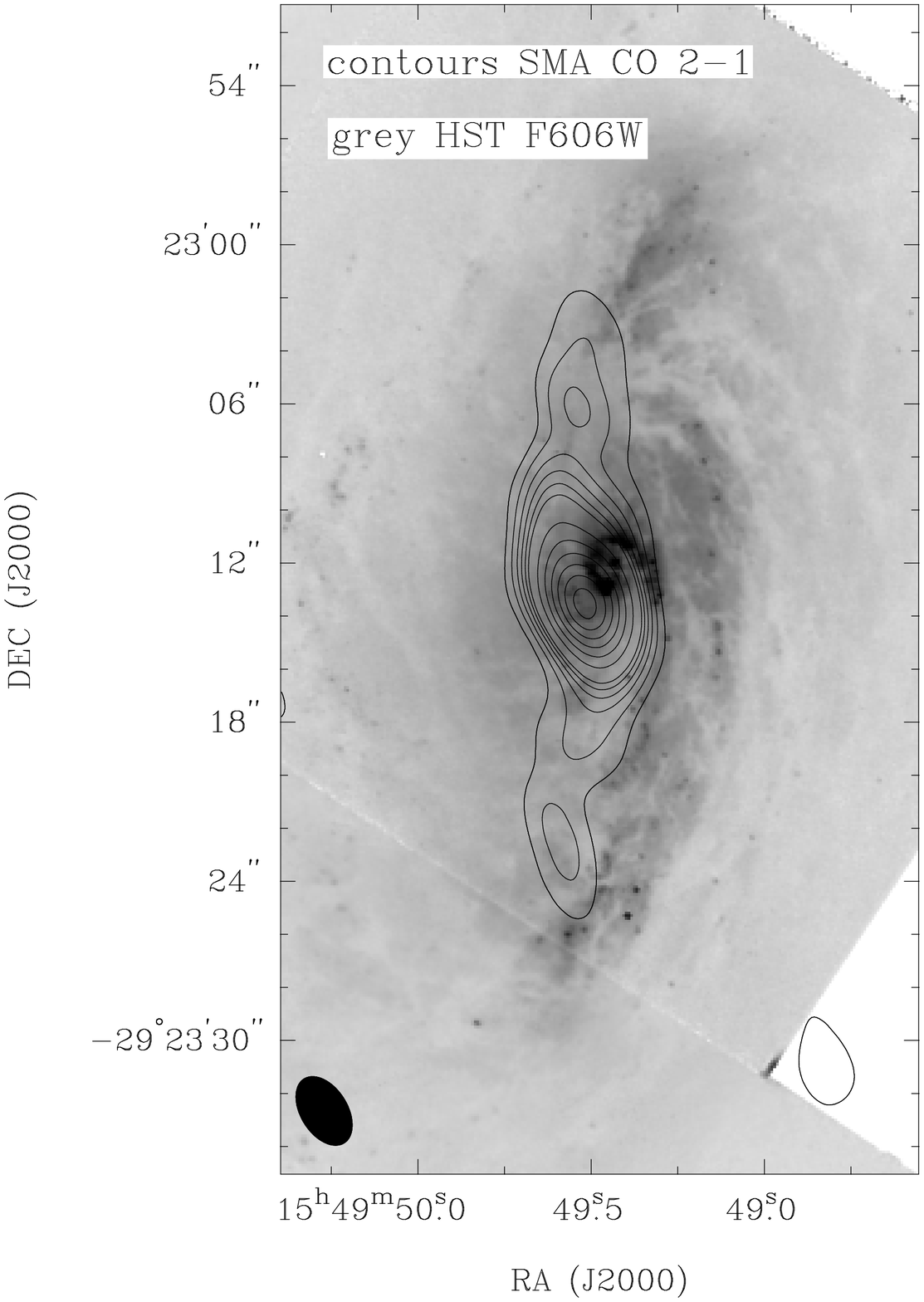}
\caption{
{\it Left panel:}
1.3\,mm continuum natural weighted emission in NGC\,6000 as measured with the SMA (contours) over an average of the J, H, and K 2MASS \citep[grey scale][]{Skrutskie06}.
Contours levels are $2\sigma$ significant as $-5$, 5, and 10\,mJy.
2MASS image is displayed in logarithmic scale.
{\it Right panel:}
Composite image of the uniform weighted $^{12}$CO $J=2-1$ integrated emission (contours) and the optical HST-WFPC2 image in the F606W filter
of NGC\,6000 \citep{carollo97}.
Contours are $3\sigma$ levels from 12 to 74\,Jy\,beam$^{-1}$\,km\,s$^{-1}$ and $10\sigma$ from 115 to 445\,Jy\,beam$^{-1}$\,km\,s$^{-1}$, where
$\sigma=4.1$\,Jy/beam\,km\,s$^{-1}$ in the image.
HST image is displayed in logarithmic scale.
The respective synthesized beams of each map are displayed in the bottom left-hand corner of each plot.
\label{fig:composite}}
\end{figure}

%: Dynamical center}
The total integrated $^{12}$CO $J=2-1$ emission detected in the nuclear region of NGC\,6000 is presented in Fig.~\ref{fig:composite} ({\it right panel}) 
on top of the optical HST image \citep{carollo97}.
In addition to the strong nuclear emission, a fainter extended structure is clearly detected ($>6\sigma$) extended in the north-south direction,
up to a distance of $\sim12''$ (1.8\,kpc) from the center.
%, likely stemming from the nuclear spiral structure.
A similar north-south structure is observed in the near-infrared (Fig.~\ref{fig:composite}, {\it left panel}).

In Fig.~\ref{fig:COmaps} we present the maps of the CO $J=2-1$ integrated emission over the range from 1900 to 2500\,km\,s$^{-1}$ ({\it Left panel})
and the intensity weighted mean velocities ({\it Right panel}).
The bright nuclear region is just barely resolved by the $2.8''\times1.7''$ ($430\times260$\,pc) beam.
A Gaussian fit to the UV visibilities in the nuclear emission results in a deconvolved extent of $3.7''\times2.1''$ ($560\times320$\,pc), centered at
$\alpha_{J2000.0}=15^{\rm h}49^{\rm m}49\fs5,\,\delta_{J2000.0}=-29^\circ23'13''$ (see Table~\ref{tab.GaussFit}).
If we assume that the deconvolved ellipticity is due to the inclination of the circumnuclear molecular disk (Sect.~\ref{Discussion}),
an inclination of $31^\circ$ is inferred. This value is in agreement with those derived from large scale images of NGC~6000.
The observed position of the CO emission peak matches the derived dynamical center and is in agreement with that derived from the continuum peak.
Similar to what is observed in the near-IR 2MASS images, the peak of CO emission does not coincide
with the optical peak of emission due to the strong extinction towards the nuclear region.
In fact, the extended CO emission follows the dust lanes seen in the optical.

\begin{figure}
\centering
\includegraphics[width=\textwidth]{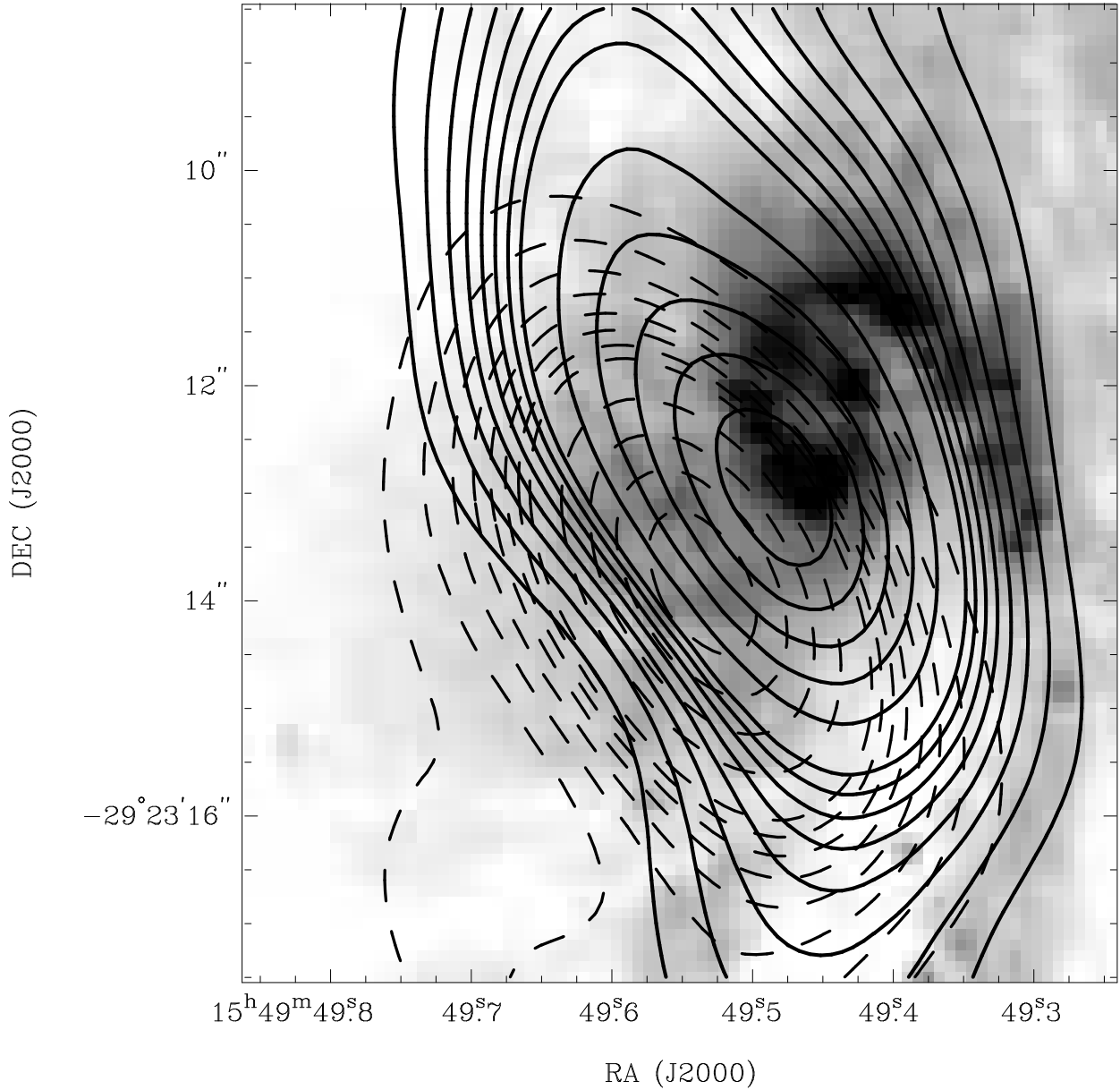}
\caption{Close-up image of the central region of NGC\,6000.
The CO red (solid contours) and blue shifted (dashed contours) emission are plotted over
the optical HST-WFPC2 image \citep{carollo97}.
Emission has been integrated $-250$ and $+250$\,km\,s$^{-1}$ with respect to the systemic velocity (2135\,km\,s$^{-1}$)
for the blue and red shifted components, respectively.
Contours are $3\sigma$ levels from 7.5 to 60\,Jy\,km\,s$^{-1}$ and $10\sigma$ from 60 to 185\,Jy\,km\,s$^{-1}$, where
$\sigma\sim2.5$\,Jy/beam\,km\,s$^{-1}$ in both images.
The optical emission is strongly asymmetrical with respect to the dynamical center.
We observe that the star forming ring, significantly offset from the center of the galaxy, is located at the position
where the highest velocity CO emission is observed as better seen in the channel maps in Fig.~\ref{fig:channelmaps}.
\label{fig:compositeZOOM}}
\end{figure}

Optical imaging of the nuclear region of NGC6000 shows 
%it as an unresolved central source with 
a bar-like structure with a nucleus surrounded by 
an irregular star forming ring with spiral arms down to the ring \citep{carollo97,carollo02}.
Fig.~\ref{fig:compositeZOOM} shows the red and blue shifted emission in a close-up view of the central region.
The star forming ring structure, offset by $(\alpha,\delta)\sim(-1''$,$1''$) from the peak of molecular emission, appears to be located 
at the peak of the high velocity CO emission (Fig.~\ref{fig:compositeZOOM}), right at the eastern edge of the molecular disk.
This structure is probably located closer to the observer than the gas disk, and therefore less affected by obscuration.
The strongly asymmetrical optical image with respect to the galactic center points to a high extinction,
as derived in Section~\ref{sec.COmass}, likely hiding the nuclear and eastern region.

%We note that the nuclear star forming-ring \citep{carollo02} is offset by $\alpha,\delta\sim(-1''$,$1''$) with respect to the peak of molecular
%emission as shown in the close up image shown in Fig.~\ref{fig:compositeZOOM}.

\subsubsection{$^{12}$CO kinematics}
\begin{figure}
\centering
\includegraphics[width=\textwidth]{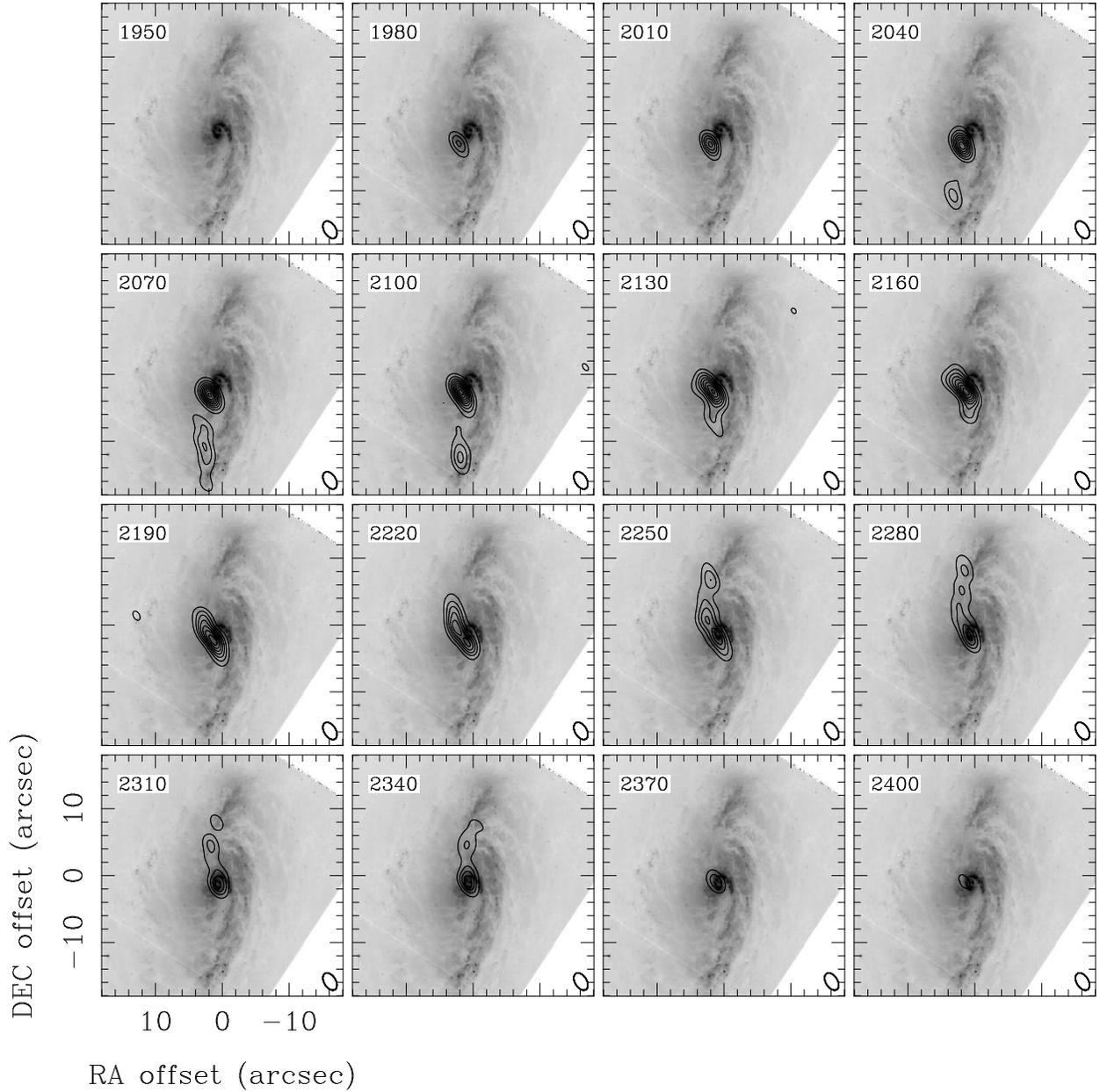}
\caption{$^{12}$CO channel maps (Contours) in steps of $\rm 30\,km\,s^{-1}$ plotted over the HST-WFPC2 image \citep{carollo97}.
Contours are plotted at $2\sigma$ levels of 1.8\,Jy\,km\,s$^{-1}$.
Central velocity of each channel is shown in the upper left corner and beam size in the lower right.
\label{fig:channelmaps}}
\end{figure}

Fig.~\ref{fig:channelmaps} shows the integrated CO emission over 30\,km\,s$^{-1}$ channels.
The resolved molecular emission in the nuclear region is observed over a range of velocities $\Delta v\rm >400\,km\,s^{-1}$ while the more extended emission
is concentrated in the central $\Delta v\rm\sim300\,km\,s^{-1}$.
As shown in the channel maps (Fig.~\ref{fig:channelmaps}) the optical bright structures are located
at the regions where the most red shifted emission ($v>2300\rm\,km\,s^{-1}$) peaks.

\begin{figure}
\centering
\includegraphics[angle=-90,width=0.4\textwidth]{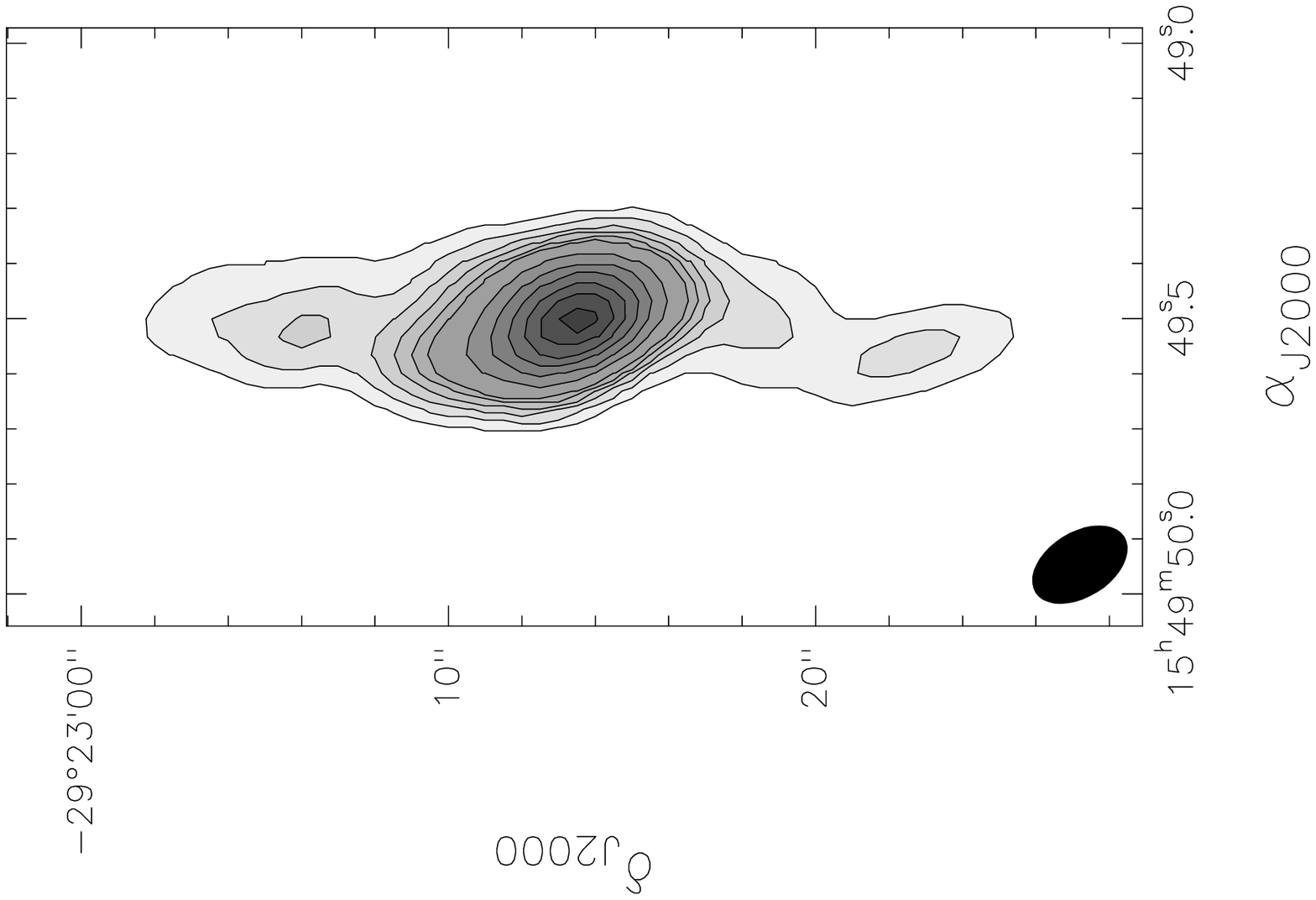}
\includegraphics[angle=-90,width=0.315\textwidth]{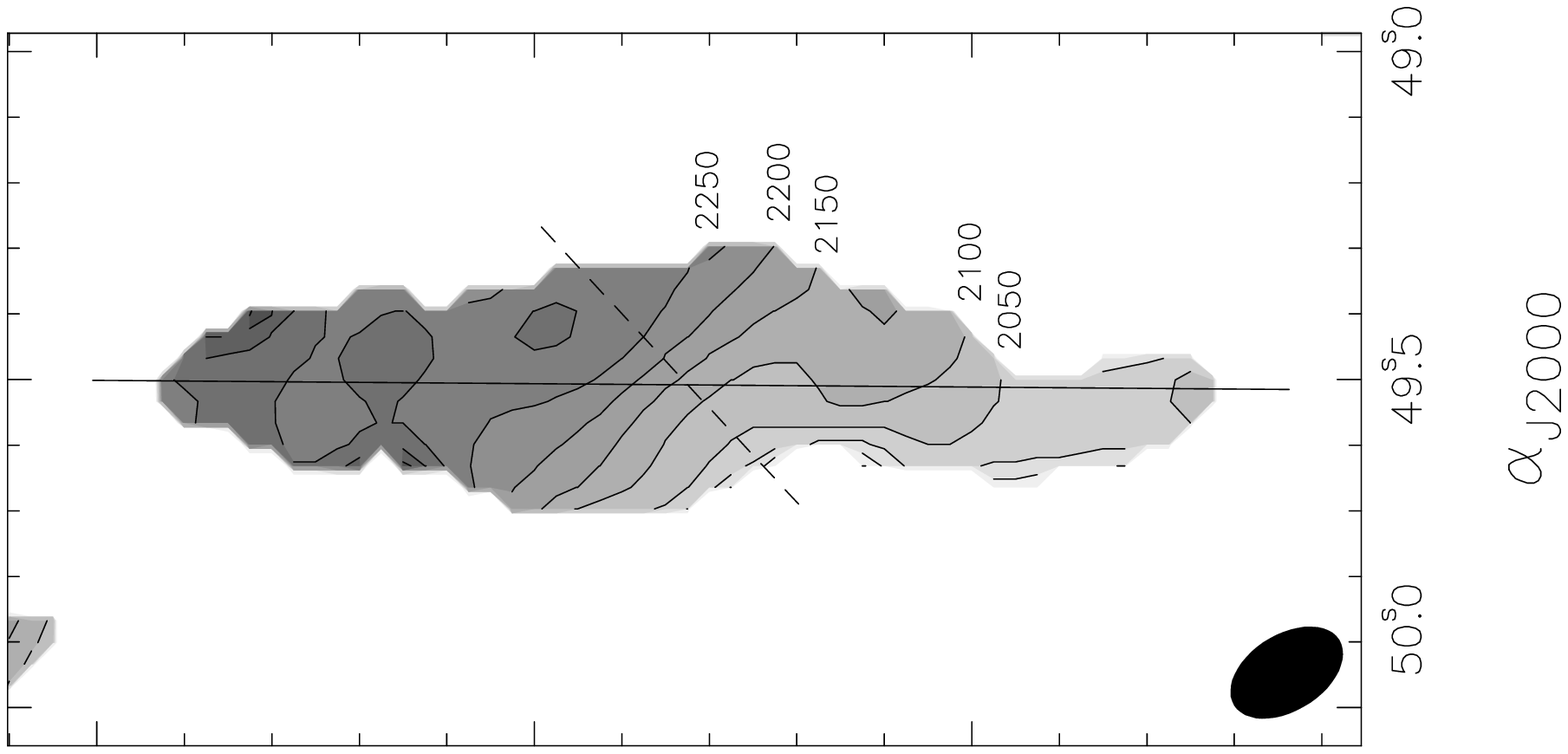}
\caption{$^{12}$CO $J=2\rightarrow1$ integrated emission and kinematics towards the nuclear region of NGC\,6000.
The synthesized beam of the uniform weighted maps ($2.8''\times1.7''$) is plotted at the bottom left-hand corner of each plot.
{\it Left panel:} Integrated flux over the 
velocity range of 1900 and 2500\,km\,s$^{-1}$.
Contours are similar to those used in Fig.~\ref{fig:composite}.
{\it Right panel:} Mean velocity field map. Contours are uniformly spaced by 50\,km\,s$^{-1}$.
Only pixel above $3\sigma$ level in integrated intensity map are shown.
The solid and dashed lines are the cut directions used for the position-velocity diagrams in Fig.~\ref{fig:pv}.
\label{fig:COmaps}}
\end{figure}

\begin{figure}
\centering
\includegraphics[angle=-90,width=0.9\textwidth]{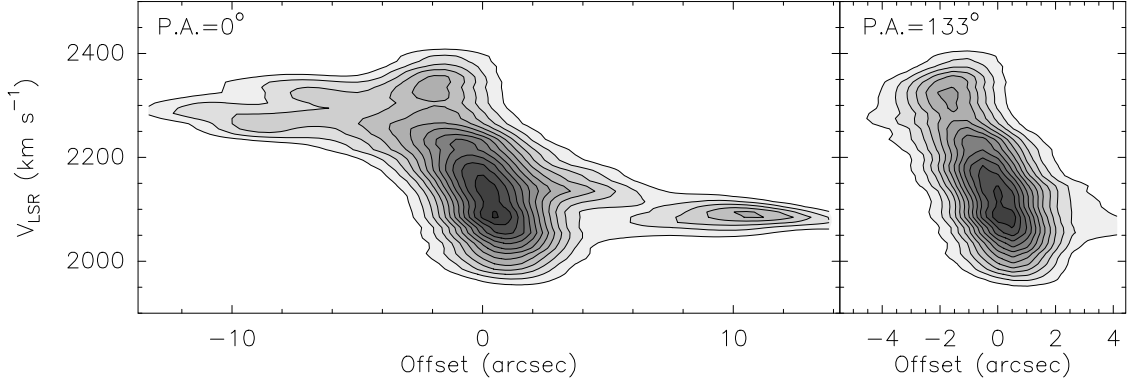}
\caption{Position-Velocity diagrams across the nuclear region of NGC\,6000 in the P.A.=$0^\circ$
and P.A.=$133^\circ$ directions.
%Left diagram has been measured in the P.A.=$0^\circ$ direction while right diagram follows 
%P.A.=$133^\circ$ 
%P.A.=$-47^\circ$ 
%direction.
Contours are $6\sigma$ levels (0.15\,Jy) from 0.15 Jy to 1.95 Jy.
Both cuts directions are shown in Fig.~\ref{fig:COmaps} by the solid and dashed lines, respectively.
\label{fig:pv}}
\end{figure}

The mean velocity map in Fig.~\ref{fig:COmaps} ({\it Right panel}) shows a strong velocity gradient in the central region,
with a smoother variation outside the nucleus.
This velocity gradient is clearly shown by the position-velocity (P-V) diagrams shown in Fig.~\ref{fig:pv}.
P-V diagrams have been plotted in two different directions,
first across the whole molecular emission in the north-south direction (P.A.=$0^\circ$; left panel in Fig.~\ref{fig:pv}) and second
in a direction perpendicular to the axis of rotation of the nuclear region 
(P.A.=$133^\circ$; right panel in Fig.~\ref{fig:pv}).
%(P.A.=$-47^\circ$; right panel in Fig.~\ref{fig:pv}).
Both directions are indicated in the right panel in Fig.~\ref{fig:COmaps} as continuous and dashed lines, respectively.
% From User Coordinates:  1961  at -2.34
% To User Coordinates:    2384  at 1.53
The P-V diagrams in both directions show how in the central $4''$ (600\,pc) the molecular gas disk is rotating like a rigid-rotor with
a rotation velocity $\sim110\rm\,km\,s^{-1}$ per arcsec, equivalent to $0.7\rm\,km\,s^{-1}\,pc^{-1}$, 
%as derived from the P.A.=$133^\circ$ and
%as derived from the P.A.=$-47^\circ$ and
after correcting the velocity for an inclination of $\sim30^\circ$ \citep{tully88,corwin85}.
This rotation velocity is similar to the average found in the sample of galaxies by \citet{Sakamoto99b}
containing barred galaxies like NGC~6000.
%The disk is strongly asymmetric towards the red shifted (south-west) side of the nucleus.

\subsubsection{$^{13}CO$ and $C^{18}O$ emission}
\begin{figure}
\centering
\includegraphics[angle=-90,width=0.7\textwidth]{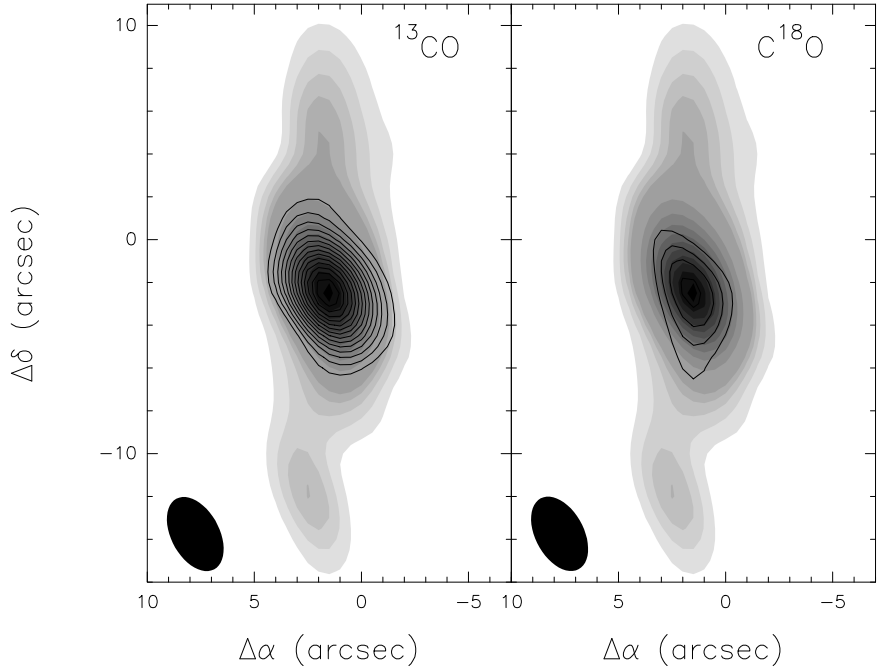}
\caption{Natural weighted integrated flux maps of the observed $J=2\rightarrow1$ transition in the different CO isotopologues (contours),
compared to the $^{12}$CO emission (grey scale).
The beam size in these maps is $3.7''\times2.2''$ (shown as filled ellipses at the bottom left corner).
$^{12}$CO grey scale levels corresponds to those in Fig.~\ref{fig:COmaps}.
Contours are $2\sigma$ levels (4.4\,Jy\,beam$^{-1}$\,km\,s$^{-1}$) in both $^{13}$CO and C$^{18}$O.
\label{fig:isotint}}
\end{figure}

\begin{figure}
\centering
\includegraphics[angle=-90,width=0.9\textwidth]{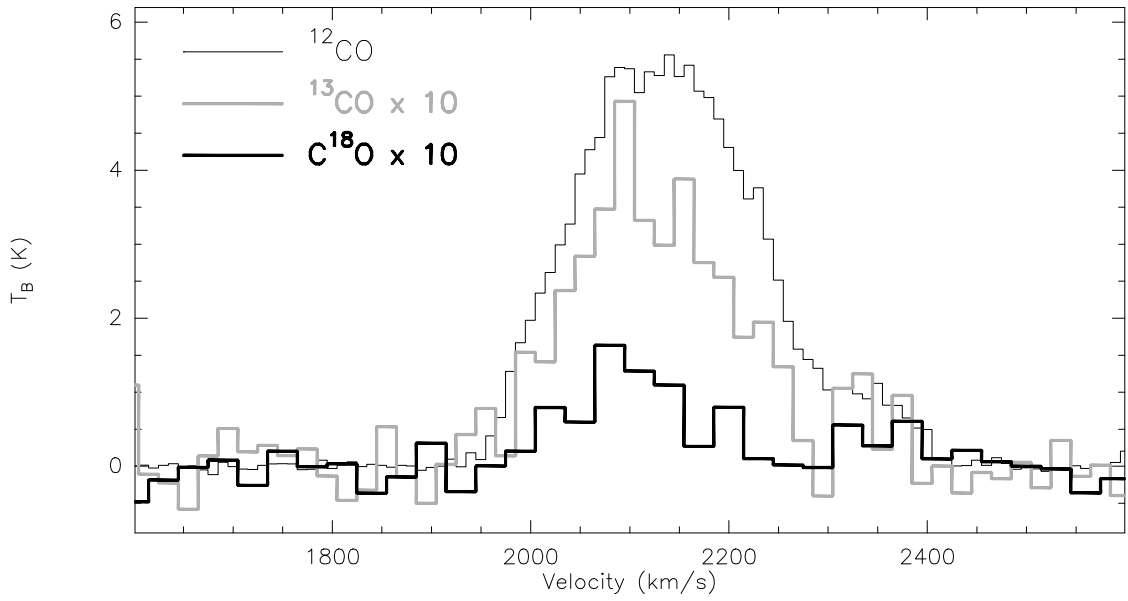}
\caption{Brightness temperature spectra of the three isotopologues extracted from the central synthesized beam of the natural weighted datacubes.
The resolution in velocity is of 10\,km\,s$^{-1}$ for the main isotope, 20\,km\,s$^{-1}$ for $^{13}$CO, and 30\,km\,s$^{-1}$ for C$^{18}$O.
The spectra of $^{13}$CO and C$^{18}$O has been multiplied by a factor of 10 for the sake of clarity in the comparison.
\label{fig:specs}}
\end{figure}

%: $^{12}$C$/^{13}$C and $^{16}$O$/^{18}$O isotopic ratios}
The emission of the rarer isotopologues were detected towards the nuclear region.
Fig.~\ref{fig:isotint} shows the integrated emission of $^{13}$CO and C$^{18}$O compared to that of the main isotopologue.
These maps have been natural weighted in order to get a higher sensitivity at the expense of a larger $3.7''\times2.2''$ ($560\times340$\,pc) beam.
The nuclear emission is resolved in both $\rm ^{13}CO$ and $\rm C^{18}O$ in the direction of the disk rotation.
However, no extended emission in the north-south direction has been detected in the rarer species due to their much lower abundance and lack
of enough sensitivity.
Still, the two lowest contours plotted for $^{13}$CO appear to be slightly elongated in the same direction as the extended emission traced by $^{12}$CO.
Gaussian models fitted to the visibilities of each isotopologue give the parameters listed in Table~\ref{tab.GaussFit}.
We observe the emission of all the three isotopical substitutions to peak at the same position.
%$\alpha_{J2000.0}=15^{\rm h}49^{\rm m}49\fs5,\,\delta_{J2000.0}=-29^\circ23'13''$.
The spectrum at the central synthesized beam of each of the natural weighted datacubes is shown in Fig.~\ref{fig:specs},
where the $^{13}$CO and C$^{18}$O spectral features have been multiplied by a factor of 10 for the sake of comparison with the much brighter CO feature.

From our observations we derive an integrated intensity ratio $R^{13}_{2-1}=I[^{12}{\rm CO}_{J=2-1}]/I[^{13}{\rm CO}_{J=2-1}]$ of $\sim 15.5$.
This ratio is in agreement with those $R^{13}_{1-0}=11.3\pm3.3$ and $R^{13}_{2-1}=12.7\pm4.7$ ratios derived from the CO $J=1-0$ and $J=2-1$
data compilations on normal galaxies with $L_{FIR}<10^{11}\,L_\odot$ \citep{Taniguchi98,Taniguchi99}.
The ratio $R^{18}_{2-1} \sim 57.9$ calculated with the $\rm C^{18}O$ isotopologue is similar to those derived for other starburst galaxies \citep{Henkel93b}.
Therefore, assuming the isotopic ratios of $^{12}$C/$^{13}$C$=40-50$ and $^{16}$O/$^{18}$O$=150-200$ apply to NGC\,6000, we can estimate an opacity of 
the main isotopologue line of $\tau_{CO (2-1)}\sim2.5-5$.
We can then accurately assume both $^{13}CO$ and $C^{18}O$ to be optically thin.

\subsection{Mass determination}
\subsubsection{Dynamical mass}
With the derived parameters in Table~\ref{tab.GaussFit} for central spectra of the CO map we can make a rough estimate
of the total dynamical mass within the nuclear region of NGC\,6000.
%As a first approximation we assume virial equilibrium as $M_{\rm dyn}=250\,M_\odot\,{\Delta v_{1/2}}^2\,R_{\rm d}$, where
%the velocity dispersion $\Delta v_{1/2}$ and the disk radius $R_{\rm d}$ are given in $km\,s^{-1}$ and $pc$, respectively \citep{Tools2004}.
%This equation implies the assumption of an spherically symmetric molecular cloud in equilibrium.
%We will consider the minor axis of the Gaussian fitted to the central CO clump as the source size,
%as it is oriented in the direction of the nuclear velocity gradient.
%Thus, the angular size of $2.1''$ is equivalent to a projected size of 322\,pc.
%To estimate the velocity dispersion we will consider the line width (FWHM), measured at the position of the CO emission peak, of 218\,km\,s$^{-1}$.
%We thus derive a total mass of $1.9\times10^9\,M_\odot$ within the central few hundred parsecs of NGC\,6000.
%This mass estimate significantly depends in the constant considered in the virial approximation which is derived from
%the assumed geometry and/or clumpiness of the molecular cloud \citep[see][for details]{Bertoldi92,Bryant96}.
The dynamical mass can be calculated as a function of the disk size and the maximum velocity corrected by inclination as
$M_{\rm dyn}=250\,M_\odot\,(V_{max}/sin(i))^2\,R_{\rm d}$, 
%in the same unit as the previous approximation.
where the $V_{max}$ and $R_{\rm d}$ are given in $km\,s^{-1}$ and $pc$, respectively.
%say '250*(1.5*153.2)*(174*2)**2'
% done with the low velocity.
For this calculation we used the peak velocity of $V_{max}\sim 180$\,km\,s$^{-1}$, with respect to the systemic velocity,
reached at a distance of $R_{\rm d}\sim1.5''$
as measured in the south-east direction of the molecular disk (low velocity end in Fig.~\ref{fig:pv}) so that we avoid the uncertainty due to the disk
asymmetry at the higher velocities (Sect.~\ref{Discussion}).
%Both approximations result in a surface H$_2$ density of $\sim5\times10^3\,M_\odot\,\rm pc^{-2}$.
This approximation
%, more accurate for a disk geometry, 
yields a total mass $M_{dyn}=7\times10^{9}\,M_\odot$.

\subsubsection{Mass of molecular gas traced by CO}
\label{sec.COmass}
% 12CO 3.75" x 2.11" beam @ 229.235
% 13CO 3.74" x 2.25" beam @ 218.781
% C18O 3.75" x 2.26" beam @ 217.949
%L'co=Tb Av Omegas Da**2
%Slambda=2kTmbOmega/lambda**2
% Xco~ 0.8 Mo (K km s-1 pc2)**-1
In order to get an accurate estimate of the mass of molecular hydrogen traced by the carbon monoxide we will use the total flux detected of
the isotopologue C$^{18}$O emission. This is equivalent to using the main isotopologue corrected by opacity effects.
Assuming optically thin emission and LTE conditions we can get the mass of hydrogen as
\begin{equation}
\label{eq.H2coldens}
N_{\rm H_2}=(7\pm2)\times10^{18}\,\frac{\rm [^{12}C^{16}O]}{\rm [^{x}C^{y}O]}\,\int{T_{\rm b}^{{\rm ^{x}C^{y}O}_{J=2-1}}\,dv}\,\,\rm cm^{-2}
\end{equation}
where $\rm ^{12}C^{16}O/^{x}C^{y}O$ is the ratio between the main form of carbon monoxide and any given isotopologue.
This expression can be applied for the three isotopologues and assumes a CO fractional abundance
$\rm ^{12}CO/H_2\sim8.5\times10^{-5}$ \citep{frerking82} and $T_{\rm ex}=30\pm10$\,K, similar to the dust temperature derived by \citet{yang07}.
Under these conditions, Eq.~\ref{eq.H2coldens} is equivalent to Eq.~3 in \citet{paglione01}.
% FLUX 33+-6 Jy km/s
% K/Jy = 2.71336
% Tdv = 90 pm 16
% NO SE DE DONDE ME SAQUE ESTO NH2 = 1.4(0.4)E+20 * 90(16)
% NH2 = 1.1(0.3)E+21 * 90(16) = 1.0 (0.3)e23
%MASA
% MALLLL TAMBIEN (383pcx199pc)*(cm/pc)**2*1.3e22*2.016g/mol/Navogadro/Msol= 1.8e7
% Cogemos el area de CO 3.7''x2.1'' = 566x321
% (566x321pc)*(cm/pc)**2*1.3e22*2.016g/mol/Navogadro/Msol= 1.8e7
% 1e23*566*321*3.086E18**2*2.016/6.0221367E23/1.99E33
With the total flux observed in C$^{18}$O and the ratio $^{16}$O/$^{18}$O$\sim150$ \citep{harrison99}, we derive an average
%$N_{\rm H_2}=(2.6\pm0.8)\times10^{22}\rm cm^{-2}$.
$N_{\rm H_2}=(1.0\pm0.3)\times10^{23}\rm cm^{-2}$.
This amount of molecular hydrogen implies a visual extinction of $A_v\sim100$ magnitudes towards the central region of NGC\,6000 which
is certainly responsible for the asymmetric optical structure of the nucleus of NGC~6000.

We estimate a total mass of molecular gas in the nuclear region of $M_{\rm H2}\sim 2.9\pm0.9\times10^8\,M_\odot$
within the central projected $\sim 566\times321$\,pc (as measured from the deconvolved CO source size).
%That is in the central $\sim 380\times200$\,pc as given by the deconvolved C$^{18}$O emission in Table~\ref{tab.GaussFit}.

From the CO map in Fig.~\ref{fig:composite} we estimate that $\sim85\%$ of the molecular mass in NGC\,6000 is concentrated in the nuclear
$\sim3''$ region ($\sim 460$\,pc).
This percentage is just a lower limit given the significant opacity affecting the CO emission mostly towards the central peak of emission.
Therefore we can estimate a total mass of molecular gas of $3.5\times10^8\,M_\odot$ within the inner $\sim1.8$\,kpc.
Assuming a correction factor 1.36 for the He and heavier elements \citep{allen73}, we can estimate a total mass of gas $M_{gas}=4.6\times10^8\,M_\odot$,
which yields a ratio $M_{gas}/M_{dyn}\sim6\%$.
%Indeed, the loss of angular momentum in the barred potential could be the cause of the high concentration of molecular gas in this galaxy.

%\subsection{Radial Distribution}
%\subsection{Rotation curve}
\section{Discussion}
\label{Discussion}

\subsection{Bar-driven molecular gas fueling the starburst in NGC~6000}
The study of the molecular gas content in a sample of 20 galaxies by \citet{Sakamoto99b} statistically shows that the gas tends to be more concentrated
in the central kiloparsec in barred systems.
This is the case of NGC\,6000 where most of the gas is located in the inner half kiloparsec
(Sect.~\ref{sec.COmass}).
The large-scale bar revealed in the NIR may also have a fingerprint in the molecular gas kinematics described in this work.
We observe that the P.A.=0$^\circ$ P-V diagram in Fig.~\ref{fig:pv} (right) shows how the extended molecular component displays a S-shaped feature.
This kinematic signature can be understood as the noncircular motions in the context of a barred potential \citep{binney92}.
The inner material could be moving in the barely resolved $x_2$ orbits which, with the resolution of our maps, would look like a molecular disk.
On the other hand the external gas would be moving in large elliptical $x_1$ orbits traced as the S-shaped profile in the P-V diagram.
Higher resolution imaging of NGC\,6000 might support the scenario of a barred potential by resolving the inner disk structure into circular $x_2$ orbits
as observed by \citet{meier08} towards Maffei\,2.
Similar signatures of a barred potential are observed in the P-V diagrams of other galaxies such as NGC\,1530 and NGC\,4258 \citep{downes96,cox96}
and equivalent structure are found in their innermost regions.

However our measured $M_{gas}/M_{dyn}$ ratio of $6\%$ is lower than the average measured by \citet{Sakamoto99b} in starbursts and barred galaxies.
This measurement is significantly affected by the way the gas mas is calculated
Indeed, adopting their same conversion factor
($X=3\times10^{19}\,\rm cm^{-2}(K\,km\,s^{-1})^{-1}$ rather than $X=0.4\times10^{19}\,\rm cm^{-2}(K\,km\,s^{-1})^{-1}$, see Sect.~\ref{COtoH2})
the measured ratio would increase up to $\sim35\%$.
This ratio is only found in barred starbursts in the \citet{Sakamoto99b} sample.

\subsection{The high velocity asymmetry}
The strong asymmetry in the nuclear ring/disk is also a particularly interesting feature.
As seen in Fig.~\ref{fig:pv} a significant part of the molecular gas traced
by CO is skewed towards the region at the lowest velocities, corresponding to the south-east disk component.
However, the emission at the highest velocities from 2300 to 2400\,km\,s$^{-1}$ in the P-V diagrams, although following the same rotation gradient,
seems to be significantly detached from the inner rotating structure.
This feature at the highest velocities corresponds to the wing observed in the spectrum shown in Fig.~\ref{fig:specs}.
We discard the possibility of self-absoption at high velocities, which would change the systemic velocity derived from the Gaussian fit to the line,
as the optically thin $\rm ^{13}CO$ and $\rm C^{18}O$ show similar profiles and are centered at the same velocity.
It is surprising that $\rm ^{13}CO$ and even $\rm C^{18}O$ are detected in this velocity component between 2300 to 2400\,km\,s$^{-1}$.
Moreover, the low ratio of $\rm ^{13}CO$ and $\rm C^{18}O$ with respect to the main isotopologue implies that a significant opacity is affecting even the
$\rm ^{13}CO$ emission in this molecular component.
CO being optically thick can provide some constraint on the extent of the emission as compared to our resolution.
From the CO brightness temperature measured $T_{\rm b}\sim1$\,K and assuming an excitation temperature $T_{\rm ex}=30$\,K 
\citep[as derived from the dust temperature,][and assuming CO being thermalized at this temperature]{yang07},
we estimate the region extent to be $<80$\,pc ($<0.5''$). This is slightly above the $\sim 50$\,pc typical size of giant molecular clouds within our Galaxy.
This giant molecular cloud within the inner region of NGC\,6000 must be very dense and more compact than our estimate to explain such CO opacity.
It could have originated as a shock between the circumnuclear disk and the infalling molecular gas along the dust lanes.
Moreover, emission arises from the same region as the star forming ring which suggest that this dense and compact molecular cloud might be directly
related to the fueling of star formation event in this region.
%the nuclear star burst event in this galaxy.

\subsection{CO-to-H$_2$ conversion factor in starburst galaxies}
\label{COtoH2}
% CON X=H2/CO=3.0e20
% CO 922 Jy km/s = 2500 pm 24
% NH2 = 7.5e23 7.2e21
We can estimate the column density of molecular gas via the conversion factor $X=N(H_2)/CO=3.0\times10^{20}\,\rm cm^{-2}(K\,km\,s^{-1})^{-1}$ based
on measurements of galactic disk molecular clouds \citep{solomon87}.
This way, and adopting the CO integrated intensity in this work, we derive a H$_2$ column density of $7.5\times10^{23}\rm cm^{-2}$ which is 
almost an order of magnitude larger than the value derived from the optically thin C$^{18}$O line (Sect.~\ref{sec.COmass}).
As already observed towards the starburst NGC~253 \citep{mauers96,harrison99}, the conversion factor in the starburst
environment might be significantly lower than that in the Galactic disk.
From our measurements we calculate a conversion factor $X=0.4\times10^{19}\,\rm cm^{-2}(K\,km\,s^{-1})^{-1}$ in agreement
with that derived for NGC\,253 \citep{mauers96}.
This result support the evidences found for a lower conversion factor in the central region of galaxies, and in particular in starburst
with respect to the Galactic disk \citep{downes98} or that measured in star forming regions in the Large Magellanic Clouds \citep{wang09}.

\section{Summary}
\label{Summary}
We have obtained high resolution ($2.8''\times1.7''$) maps of the molecular emission towards NGC\,6000 as traced by the $J=2-1$ transition
of carbon monoxide with the SMA.
Emission from the three main isotopologues of carbon monoxide, $^{12}$CO, $^{13}$CO, and C$^{18}$O are detected.
Both the continuum and molecular emission peak at the same position.
From these measurements we have accurately determined the nucleus position in NGC\,6000 at
$\alpha_{J2000.0}=15^h49^m49\fs5$ and $\delta_{J2000.0}=-29\arcdeg23\arcmin13$, in agreement with previous IR determinations.
More than an $85\%$ of the emission is concentrated towards the nuclear projected $\sim 400\times300$\,pc region, where we estimate
optical extinctions of the order of $\sim100$ magnitudes.
Such large obscuration is likely responsible for the asymmetry in the optical images, probably hiding the eastern part of the nuclear region at
these wavelengths.

The emission from the central region is resolved in the direction of the observed velocity gradient.
Molecular emission is detected in the north-south direction extended up to distances of $>1.5$\,kpc from the center.
However this is only detected in the main isotopologue.
Both the nuclear and extended emission show dynamical evidences that might be explained by the presence of a bar potential affecting the molecular material,
likely responsible of the strong nuclear concentration and maybe the starburst event.
We observed a high velocity component which appears to be detached from the nuclear disk/ring and is highly optically thick.
With an estimated extent of $<80$\,pc, this giant molecular cloud is likely to be related to the fueling of the nuclear star burst in NGC\,6000.

%Isotopologues line ratios are similar to those observed in other starburst galaxies.
%We used a non-LTE approximation to derive the isotopic ratios from the observations of the $^{13}\rm CO$ and $\rm C^{18}O$ isotopologues.
%The estimated isotopic ratios are $^{12}$C/$^{13}$C=$20\pm1$ and $^{16}$O/$^{18}$=$76\pm14$, which are a factor of two lower than the values found
%in other starburst galaxies. This difference is likely due to an underestimation of the CO opacity. Assuming a larger opacity yields
%ratios similar to those observed in other starbursts galaxies. 
%Indeed the observed $^{12}$CO/$^{13}$CO line ratio is similar to that found in other galaxy nuclei.
The ratios of $^{13}$CO and C$^{18}$O with respect to the main isotopologue are similar to the average measured in similar starburst galaxies with luminosities
$L_{FIR}<10^{11}L_\odot$ in the literature \citep{Henkel93b}.

The total dynamical mass derived for NGC\,6000 is $1.9\times10^{9}\,M_\odot$.
% out of which an $\lesssim1\%$ is estimated to be in the form of molecular gas.
We estimate a total mass of molecular gas of $4.6\times10^8M_\odot$ and a ratio $M_{gas}/M_{dyn}\sim6\%$, similar to the starbursts
in the barred galaxy sample by \citep{Sakamoto99b}.
The independent estimate of the molecular mass with the optically thin C$^{18}$O isotope allows the determination of the conversion
factor CO-to-H2 of $X_{CO}=0.4\times10^{19}\,\rm cm^{-2}(K\,km\,s^{-1})^{-1}$, which agrees with previous estimates that measure a lower factor in
starburst nuclear regions \citep{mauers96} as compared with the standard factor derived for Galactic disk clouds.

\acknowledgments
This paper uses data taken with the Submillimeter Array on Director's discretionary time as part of the Astronomy 191 course for
Harvard University undergraduates.
We want to thank Prof. Carl Heiles for his valuable comments on the manuscript.
The Submillimeter Array is a joint project between the Smithsonian Astrophysical Observatory and the Academia Sinica Institute of Astronomy 
and Astrophysics and is funded by the Smithsonian Institution and the Academia Sinica.

{\it Facilities:} \facility{SMA ()}

\newpage

\begin{deluxetable}{lcc}
\tablecaption{Summary of observational parameters for NGC 6000\label{galaxyparams}}
\tabletypesize{\footnotesize}
\tablewidth{0pt}
\tablehead{\colhead{Parameter} & \colhead{Value} & \colhead{Reference}}
\startdata
\multicolumn{3}{c}{\it GALAXY POSITION}\\
%CO $2-1$                &               15:49:49.5, -29:23:13.5         &     This work \\
ROSAT ($J2000$)         &               15:49:50.1, -29:23:10.7         &       1       \\
IRAS  ($J2000$)         &               15:49:49.4, -29:23:11           &       2       \\
$V_{\rm LSR}$           &               $2145$\,km\,s$^{-1}$            &       3       \\
Distance ($H_0=75\,km\,s^{-1}\,Mpc^{-1}$) &  31.6\,Mpc                  &       4       \\
\multicolumn{3}{c}{\it MORPHOLOGY}\\
%Morphological type      &               SB(s)bc=SBS4*, Scd              &       5,6     \\
Morphological type      &               SB(s)bc              &       5     \\
Inclination             &               33\degr, 27\degr                &       6,7     \\
$\log\,D_{25}\,(0.1\arcmin)$    &             1.27                            &       5       \\
%$\log(R_{25})$          &               0.06                            &       5       \\
Weighted mean luminosity class  &        4.0                             &       7       \\
$Diameter_{maj\times min}$      &     2.3\arcmin $\times$ 2.0\arcmin    &    8          \\
P.A.                            &      154\degr,163\degr                &    5,9        \\
%P.A. (East of North, radio)     &      154\degr,163\degr                &    5,9        \\
\multicolumn{3}{c}{\it PHYSICAL PARAMETERS}\\
T$_{dust}$\,(K)                 &               34, 30                  &     10,11     \\% 	10, 16, 32	\\
$M_{dust}\,(M_\odot)$           & $1.5 \times 10^9$, $2.4\times 10^7$   &     10,11     \\
$\log\,L_{FIR}\,(\,L_{\odot})$  &               $10.83$                 &     12        \\
$\log\,L_{IR}\, (\,L_{\odot})$  &               $10.97$                 &     12        \\
$\log\,M_{HI}\,(M_{\odot})$     &               $9.86,\, 9.8$           &     13,14     \\
$\log\,M_{H2}\,(M_{\odot})$     &               $8.5$                   &     15        \\
$\log\,M_{gas}\,(M_{\odot})$    &               $8.7,10.0$              &     15,16     \\
$\log\,M_{dyn}\,(M_{\odot})$    &               $9.8$                   &     15     \\
\enddata
\tablerefs{
(1) \citealt{boller92};
(2) \citealt{sanders03};
(3) \citealt{albrecht07};
(4) \citealt{pizzella05};
(5) \citealt{RC3};
(6) \citealt{tully88};
(7) \citealt{corwin85};
(8) \citealt{dressler91};
(9) \citealt{young95};
(10) \citealt{roche93};
(11) \citealt{yang07};
(12) \citealt{sanders03};
(13) \citealt{martin91};
(14) \citealt{koribalski04};
(15) This work;
(16) \citealt{chini95};
}
%apj
%\end{deluxetable*}
%manuscript
\end{deluxetable}

\newpage 

\begin{deluxetable}{l c c c}
\tablecaption{Derived parameters from the CO, $^{13}$CO and C$^{18}$O observed emission \label{tab.GaussFit}}
\tablewidth{0pt}
\tablehead{\colhead{Parameter} & \colhead{CO}                         & \colhead{$^{13}$CO}                   & \colhead{C$^{18}$O}} 
\startdata
\multicolumn{4}{c}{\it Gaussian fit to the maps \tablenotemark{a}} \\
%Center ($J2000.0$)             & 15:49:49.5, -29:23:13.3              & 15:49:49.5, -29:23:13.3               & 15:49:49.5, -29:23:13.1        \\
Center ($J2000.0$)             & 15:49:49.5, -29:23:13                & 15:49:49.5, -29:23:13                 & 15:49:49.5, -29:23:13            \\
Size (FWHM)                    & $3.7''\times2.1''$, P.A. $12^\circ$  & $2.7''\times2.1''$, P.A. $29^\circ$   & $2.5''\times1.3''$, P.A. $5^\circ$  \\
Total Flux                     & $922\pm9$\, Jy\,km\,s$^{-1}$         & $90\pm7$\,Jy\,km\,s$^{-1}$            & $33\pm6$\,Jy\,km\,s$^{-1}$              \\
%Total Flux                     & 1.0809665845907E+03                  &
\multicolumn{4}{c}{\it Central pixel spectra \tablenotemark{b}} \\
%$V_{LSR}$                      & $2135.3\pm0.2$ km\,s$^{-1}$          & $2119\pm4$ km\,s$^{-1}$               & $2105\pm9$ km\,s$^{-1}$             \\
%$\Delta v_{1/2}$               & $217.45\pm0.01$ km\,s$^{-1}$         & $196\pm10$ km\,s$^{-1}$               & $150\pm20$ km\,s$^{-1}$             \\
%$T_{\rm B,\, peak}$            & 157 mK                               & 10 mK                                 & 4 mK                               \\
$V_{LSR}$                      & $2135.3\pm0.3$ km\,s$^{-1}$          & $2119\pm4$ km\,s$^{-1}$               & $2104\pm9$ km\,s$^{-1}$             \\
$\Delta v_{1/2}$               & $217.1\pm0.6$ km\,s$^{-1}$           & $196\pm10$ km\,s$^{-1}$               & $150\pm20$ km\,s$^{-1}$             \\
$T_{\rm B}$                    & 5600\,mK                             & 400\,mK                               & 140\,mK                             \\
\enddata
\tablenotetext{a}{Results of the deconvolved Gaussian fit models applied to the UV visibilities.}
\tablenotetext{b}{Parameters from the Gaussian fit to the spectra at the central pixel of each map.}
\end{deluxetable}

\newpage
\bibliographystyle{apj}	% see astronat package, apj.bst
\bibliography{ngc6000.bib}	% links to ngc6000.bib for bibtex information

\end{document}